\pdfoutput=1
\documentclass[a4paper]{article}

\usepackage{icrc2013}
\usepackage[english]{babel}
\usepackage[latin1]{inputenc}
\usepackage{booktabs}
\usepackage{units}
\usepackage{color}
\usepackage{amsmath,amssymb}
\usepackage[mathlines,switch]{lineno}
\usepackage{ifthen}
\usepackage[english]{babel}

\newcommand{\pcmode}{0}

\ifthenelse{\equal{\pcmode}{1}}{
  \usepackage{graphicx}
  \usepackage{mathptmx}
  \usepackage[pdftex,colorlinks=true]{hyperref}
  \usepackage{setspace}
  \doublespacing
  \linenumbers
}
{
\usepackage{icrc2013}
}

\newcommand*\patchAmsMathEnvironmentForLineno[1]{%
  \expandafter\let\csname old#1\expandafter\endcsname\csname #1\endcsname
  \expandafter\let\csname oldend#1\expandafter\endcsname\csname end#1\endcsname
  \renewenvironment{#1}%
     {\linenomath\csname old#1\endcsname}%
     {\csname oldend#1\endcsname\endlinenomath}}%

\makeatletter
\DeclareRobustCommand*{\bfseries}{%
  \not@math@alphabet\bfseries\mathbf
  \fontseries\bfdefault\selectfont
  \boldmath
}
\makeatother

\def\scix#1#2{#1 \! \times \! 10^{#2}}

\newcommand{\eV}{\ensuremath{\text{e\kern-0.07em V}}}

\title{Highlights from the Pierre Auger Observatory}

\shorttitle{\includegraphics[width=2cm]{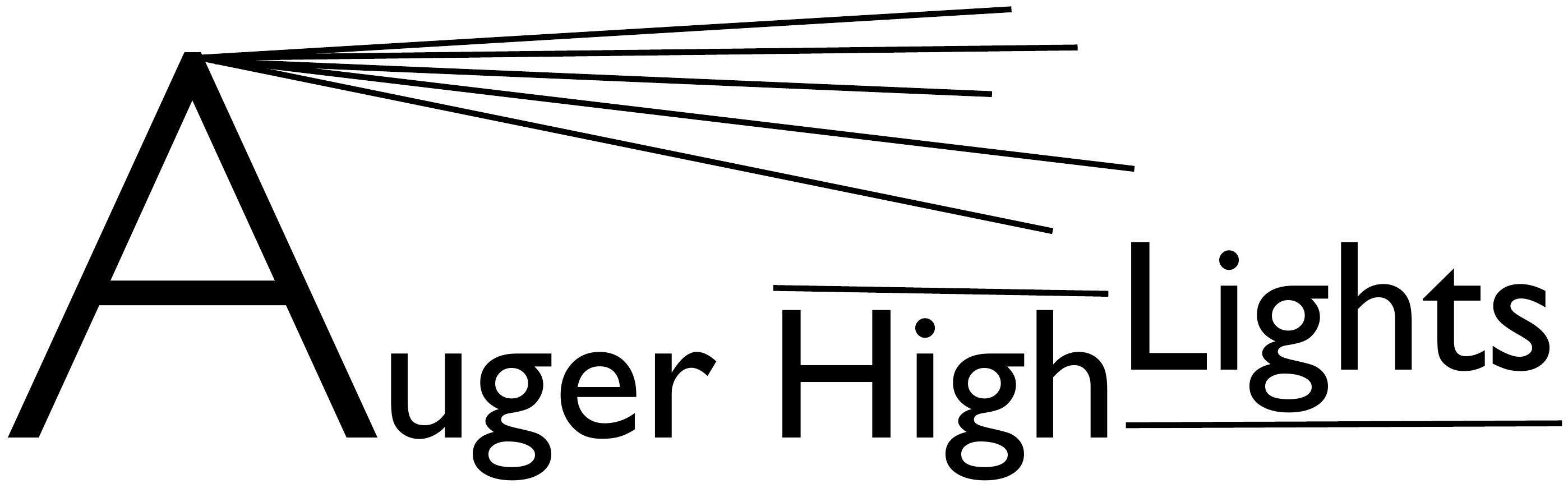}}

\authors{Antoine Letessier-Selvon$^{1}$ for the Pierre Auger Collaboration$^{2}$.
}

\afiliations{
$^1$ Laboratoire de Physique Nucléaire et des Hautes Énergies, Université Pierre et Marie Curie et Université Denis Diderot, CNRS/IN2P3, Paris, France.\\
$^2$ Full author list: http://www.auger.org/archive/authors 2013 05.html \\
}

\email{auger\_spokespersons@fnal.gov}

\abstract{The Pierre Auger Observatory is the world's largest cosmic ray
observatory. Our current exposure reaches nearly 40,000 km$^2$ str and
provides us with an unprecedented quality data set. The performance and
stability of the detectors and their enhancements are described. Data
analyses have led to a number of major breakthroughs. Among these we
discuss the energy spectrum and the searches for large-scale anisotropies. 
We present analyses of our \Xmax\, data and show how it
can be interpreted in terms of mass composition. We also describe some new
analyses that extract mass sensitive parameters from the 100\% duty cycle
SD data.  A coherent interpretation of all these recent results opens
new directions. The consequences regarding the cosmic ray composition
and the properties of UHECR sources are briefly discussed.}


\begin{document}
\newcommand{\Xmax}{X$_{\rm max}$}
\newcommand{\km}{km$^2$}

\maketitle

\section{The Pierre Auger Observatory}
The Pierre Auger Collaboration is composed of more than 500 members from 19 different countries. The observatory~\cite{Auger}, the world's largest,
is located in the southern part of the province of Mendoza in Argentina. It is dedicated to the studies of Ultra High Energy Cosmic Rays (UHECR) from a fraction
 of EeV\footnote{1 EeV = 10$^{18}$ eV or 0.16 J} to the highest energies ever observed at several hundreds of EeV.  
The Observatory comprises several instruments working in symbiosis :
\begin{itemize}   
\item A surface detector array (SD) of 1600 water Cherenkov detectors (WCD) arranged on a regular triangular grid of 1500 m and covering 3000 \km~\cite{AugerSD}.
\item 4 sites with fluorescence detector (FD)  (each site contains 6 telescopes for a total of 180$^0$ azimuth by 30$^0$ zenith field of view)~\cite{AugerFD}.
\item A subarray, the Infill,  with 71 water Cherenkov detectors on a denser grid of 750 m covering nearly 30 \km~\cite{AugerInfill}.  This subarray is part of the AMIGA extension that will also have buried muon counters at each 71 WCD locations (7 are in place~\cite{AC:Federico}). 
\item 3 High Elevation Auger Telescopes (HEAT) located at one of the fluorescence site~\cite{HEAT} dedicated to the fluorescence observation of lower energy showers.
\item A subarray of 124 radio sensors (AERA, Auger Engineering Radio Array) working in the MHz range and covering 6\km~\cite{AERA}.
\item A sub Array of 61 radio sensors (EASIER, Extensive Air Shower Identification with Electron Radiometer) working in the GHz range and covering 100\km~\cite{EASIER}.
\item Two GHz imaging radio telescope AMBER~\cite{AMBER} and MIDAS~\cite{MIDAS} with respectively 14ºx14º and 10ºx20º  field of views. 
\end{itemize}
The three last items are R\&D on the detection of extensive air showers using the radio emission of the EM cascade in the atmosphere.

In total the Auger collaboration has provided to this conference  32 contributions\,\cite{AugerContrib},  including 3 contributions~\cite{Auger-TA-1,Auger-TA-2,Auger-TA-3}
done in collaboration with the Telescope Array collaboration (TA)~\cite{TA}. These contributions describe the wide range of  detector techniques, analyses tools, monitoring system and scientific results
developed and produced by the collaboration. In this short highlight only a fraction of those contributions can be presented. 

After a brief description of the detector status and of the data selection, we 
present the updated energy scale and corresponding energy spectra, as measured by the various components of the observatory. We also report on the measurements of the two first moments (mean and variance) 
of the longitudinal shower profile \Xmax\,distributions in several energy bins and interpret them in terms of mass composition using recent update of the high energy generators~\cite{EPOS-LHC,QGSJetII-04}.
 \begin{figure}[!t]
  \includegraphics[width=\columnwidth]{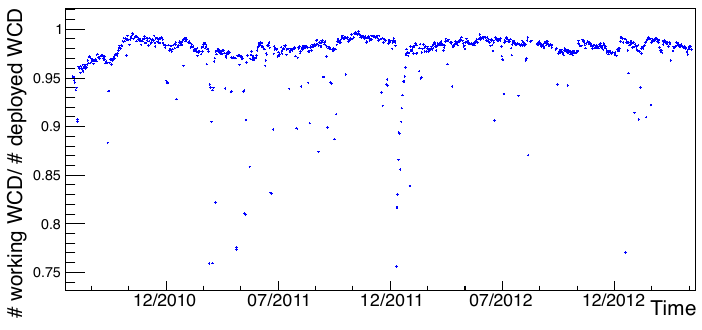}
  \caption{ Normalized number of active SD stations as a function of time (\cite{AC:Carla}).}
  \label{fig:moni1}
 \end{figure}

We describe new analysis techniques that allow us to measure the muonic content of extensive air showers. The analyses, based on the SD data set, profit from the high statistics from this detector with nearly 100\% duty cycle. They allow us to confront  models for hadronic interactions at high energies with data at the highest energies and also to recover mass sensitive parameters independently from the FD measurements. 

Last but not least we report on the searches for large scale anisotropies in the EeV range, and their consequences.

\begin{table*}[!t]
\centering
\begin{tabular}{lllll}
\toprule
 & \multicolumn{3}{c}{{\bf Auger SD}} & \multicolumn{1}{c}{{\bf Auger hybrid}} \\
\cmidrule(r){2-4}
 & {\bf \unit[1500]{m} vertical} & {\bf \unit[1500]{m} inclined} & {\bf \unit[750]{m} vertical} & \\
\midrule
Data taking period & 01/2004 - 12/2012 & 01/2004 - 12/2012 & 08/2008 - 12/2012 & 11/2005 - 12/2012 \\
Exposure $\left[ \mathrm{km^2 ~ sr ~ yr} \right] $ & $31645 \pm 950$ & $8027 \pm 240$ & $79 \pm 4$ & -\\
Zenith angles~$\left[{}^\circ\right]$ & $0-60$ & $62-80$ & $0-55$ & $0-60$ \\
Threshold energy $E_\mathrm{eff}$~$\left[\unit{\eV}\right]$ & $\scix{3}{18}$ & $\scix{4}{18}$ & $\scix{3}{17}$ & $10^{18}$ \\
No. of events ($E>E_\mathrm{eff}$) & 82318 & 11074 & 29585 & 11155 \\
No. of events (golden hybrids) & 1475 & 175 & 414 & -\\
Energy calibration (A)~$\left[\unit{E\eV}\right]$ & $0.190 \pm 0.005$ & $5.61 \pm 0.1$ & $(1.21 \pm 0.07) \cdot 10^{-2}$ & -\\
Energy calibration (B) & $1.025 \pm 0.007$ & $ 0.985 \pm 0.02 $ & $1.03 \pm 0.02$ & -\\
\bottomrule
\end{tabular}
\caption{\small Summary of the experimental parameters describing data of the different measurements at the Pierre Auger Observatory. Numbers of events are given above the energies corresponding to full trigger efficiency (\cite{AC:Alexander}).}
\label{tb:datasets}
\end{table*}

  \begin{figure}[!t]
  \includegraphics[width=\columnwidth]{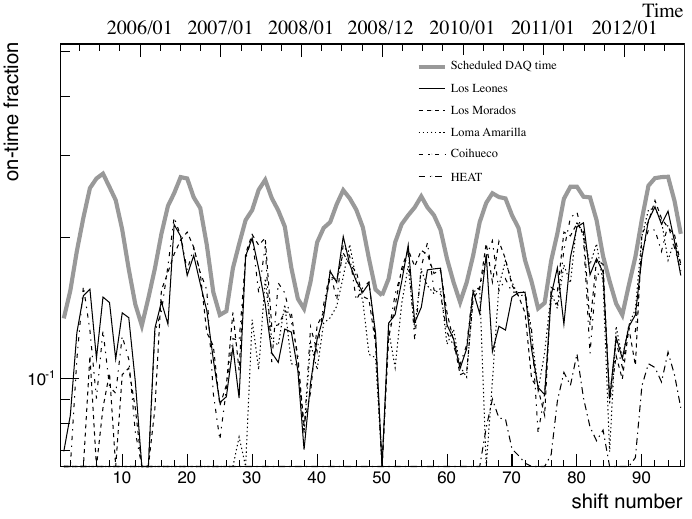}
  \caption{ Hybrid on-time fraction for the four FD sites and HEAT. The thick gray line defines the scheduled data-taking (limited to nights with less than 60\% moon-fraction. (\cite{AC:Carla}).}
  \label{fig:moni2}
 \end{figure}
\subsection{Status}
The hybrid concept has been pioneered by the Auger collaboration and allows, among other things,  for calibration of the SD that is fully data driven, thus avoiding the uncertainties related to the use on Monte Carlo simulated showers.  Such calibration allows the transfer of the high precision calorimetric information collected by the FD to the 100\% duty cycle SD. In the following the term {\it hybrid} will also refer to those events that are observed simultaneously by the SD and FD, they form a specific data set called the  hybrid data. 

To fully benefit from this technique it is however mandatory to monitor with extreme precision both the detectors activity and the atmospheric experimental conditions. Out of the major correction terms applied to the FD energy, the atmospheric transmission through aerosols has the largest time variation and must be followed most closely. 

The Auger site is equipped with an extensive set of instruments that measure the atmospheric conditions~\cite{AC:Laura,AC:Maria,AC:Johana}. These instruments allow us to determine within accuracies of a few percent the hourly vertical aerosol optical depth (VAOD) as well as to obtain a sky representation of the cloud coverage.

In addition to the atmosphere monitoring, an extensive  collection of hardware and software tools have been developed and are used to monitor (up to second by second) the activity of the different detector components. This provides on-line as well as long term detector and data quality control~\cite{AC:Carla}. Examples of such monitoring information are shown in figure~\ref{fig:moni1} and~\ref{fig:moni2}. 

In Fig.~\ref{fig:moni1} the activity of each individual WCD station is reported (the data averaged in the plot is collected each second). One can visually measure the nearly constant and efficient activity of the array which is about 98\% on average. 

In Fig.~\ref{fig:moni2} we show the hybrid on-time fraction of our FD sites. Such monitoring allows for a precise determination of the experimental exposure as well as for a precise control of the data quality.

\subsection{Data sets}
The data sets used for the various analyses presented here and at the conference have minor variations from one analysis to the next as described in detail in the corresponding conference contributions~\cite{AugerContrib}. However,  they share some common features. 

The data taking period extends from 1 January 2004 to 31 December 2012, thus updating the measurements we have published earlier. To ensure an appropriate and accurate reconstruction of the cosmic ray parameters such has the arrival direction and energy or of the characteristics of the shower longitudinal development (e.g. \Xmax) several quality cuts are applied. For the SD analyses it is for example required that the WCD with the largest signal be surrounded by six working and active WCDs at the time of the event. 

Different attenuation characteristics of the electromagnetic and muonic shower components lead to different reconstruction methods for the different associated zenith angle ranges. We distinguish in particular between \textit{vertical events} with a zenith angle $\theta$ between 0 and  $60^\circ$ (or $\theta< 55^\circ$ for the Infill) and \textit{inclined events} with a zenith angle between $62$ and  $80^\circ$. 

As mentioned, the energies of SD events are determined from the cross calibration with the FD using the hybrid data set. The SD size parameters $S$ ($S_{38}$, $S_{35}$ and $N_{19}$), for the regular array, the 750~m Infill and the inclined data sets respectively, are related to the FD energy using a calibration curve of the form $E_{FD} = A\,S^B$. The value of those parameters are reported in Table~\ref{tb:datasets}\,together with the corresponding data sets sizes and main characteristics. 

The overall up-time and efficiency of the SD is about 98\% while that of the FD is 13\%. The  energy resolution of the SD alone is 12\% (statistical) above 10 EeV while 
the angular resolution is less than 1$^\circ$ in that energy range.  

The total exposure, corresponding to the data sets presented in table~\ref{tb:datasets} is about 40,000 \km\,sr\,yr. From now on, over 6~000\km\,sr\,yr are expected to be collected each year. 

It is interesting to note that the combination of our horizontal and vertical data sets gives us a remarkably large sky coverage (up to nearly 50$^\circ$ declination North). In addition, a recent upgrade of our triggering 
system, especially at the local WCD level, is being commissioned. It will allow us to bring the energy at which the SD reaches full trigger efficiency from 3~EeV down to about 1~EeV and to significantly improve our photon sensitivities in the EeV range.  
\begin{table}[!t]
\begin{center}
\begin{tabular}{|c|c|c|}\hline 
                                                                      \multicolumn{2}{|c|}{\bf Changes in FD energies at ${\bf 10^{18}}$ eV}  \\ \hline
Absolute fluorescence yield                     &  {\bf -8.2\%}               \\ \hline
New optical efficiency                                                   &         4.3\%               \\ 
Calibr. database update                                                &         3.5\%              \\ 
Sub total (FD calibration)        &   {\bf 7.8\%}               \\ \hline
Likelihood fit of the profile                                            &   2.2\%               \\ 
Folding with the point spread function                           &  9.4\%              \\ 
Sub total (FD profile reconstruction )     &   {\bf 11.6\% }              \\ \hline 
New invisible energy                   &    {\bf 4.4\%}               \\ \hline
 Total                                                                               &   {\bf 15.6\%}             \\ \hline
\end{tabular}
\end{center}
\caption{Changes to the shower energy at $10^{18}$ eV (\cite{AC:Valerio}).}
\label{tab:energy_shifts}
\end{table}

\begin{figure}[!t]
\includegraphics[width=\columnwidth]{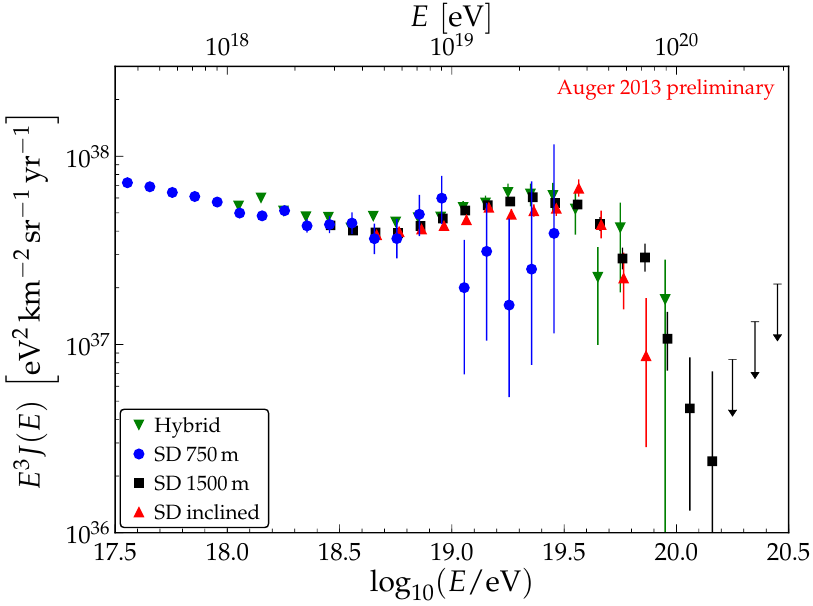}
\caption{The Auger energy spectra obtained from the various SD and hybrid data sets. (\cite{AC:Alexander}).}
\label{spectra}
\end{figure}

\subsection{Absolute Energy Scale}
On top of the extensive monitoring of the atmosphere and of the FD operation as a function of time, one must also perform very detailed studies of the light collection efficiencies, and frequently calibrate or check the calibration of the instruments. An extensive campaign of measurements and control have been performed at Auger to improve the knowledge of our energy scale and to reduce the systematic uncertainties associated
with it~\cite{AC:Valerio}.  

Corrections to the absolute energy scale come from various sources. Among these are the fluorescence yield~\cite{AIRFLY},  the point spread function measurements performed with our flying light source (the Octocopter now also jointly used at TA~\cite{Auger-TA-1}), the changes in the reconstruction of the shower longitudinal profile, the better understanding of the telescope point spread function and accurate simulation of the optics through detailed ray-tracing~\cite{AC:Julia}, the improvements in the analyses and in particular in the estimation of the missing energy~\cite{AC:Matias}.  A summary of the changes at a reference energy of 1 EeV is given in table~\ref{tab:energy_shifts}, amounting to +15.6\%.  There is an small energy dependence associated with some of those corrections and the global shift becomes +11.3\% at 10 EeV.  

These extensive studies also have allowed better control of the uncertainties associated with each of those corrections. While our overall systematic uncertainty was 22\% at the 32nd ICRC in Beijing (China, 2011), it is now reduced to 14\%.
\begin{figure}[!t]
\includegraphics[width=\columnwidth]{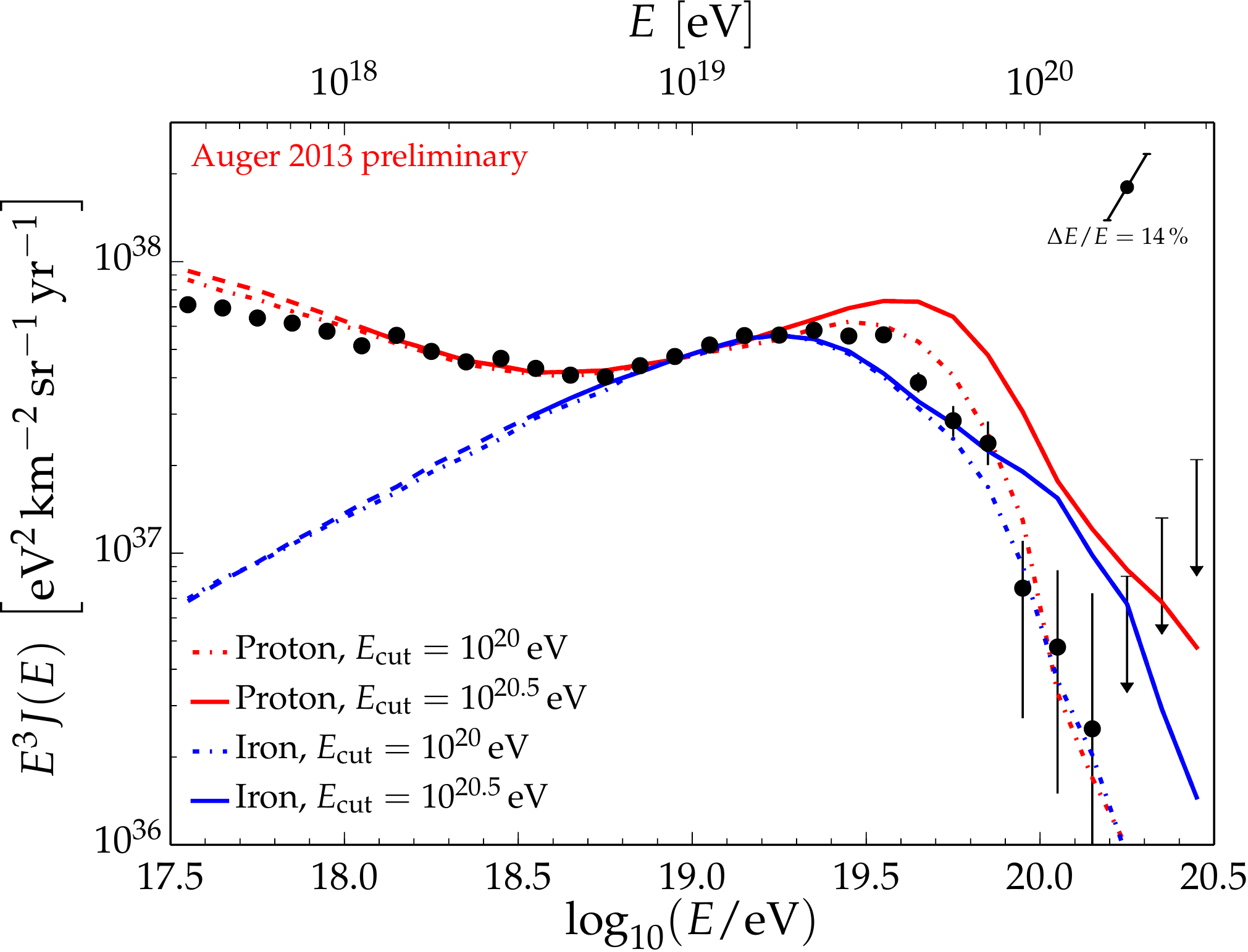}
\caption{The combined Auger energy spectrum compared to spectra from different astrophysical scenarios.}
\label{f:comb_astromodels}
\end{figure}

\section{Spectrum}
After energy calibration the exposure for each  data set  (hybrid, Infill, SD vertical and SD horizontal)  is  carefully evaluated on the basis of our precise monitoring systems. The corresponding spectra are shown in Fig.~\ref{spectra}. 

Those spectra are combined to form the Auger spectrum as shown in figure~\ref{f:comb_astromodels}.  The combination process relies upon a maximum likelihood method that allows for a normalization adjustment between the various spectra~\cite{AC:Alexander}. The corrections, which are  well within the normalization uncertainty of the individual spectra, amount to -6\%, +2\%, -1\% and +4\% respectively. The total number of events comprising the spectrum shown in figure~\ref{f:comb_astromodels} is about 130,000.

\begin{figure*}[!t]
\center
\includegraphics[width=0.48\textwidth]{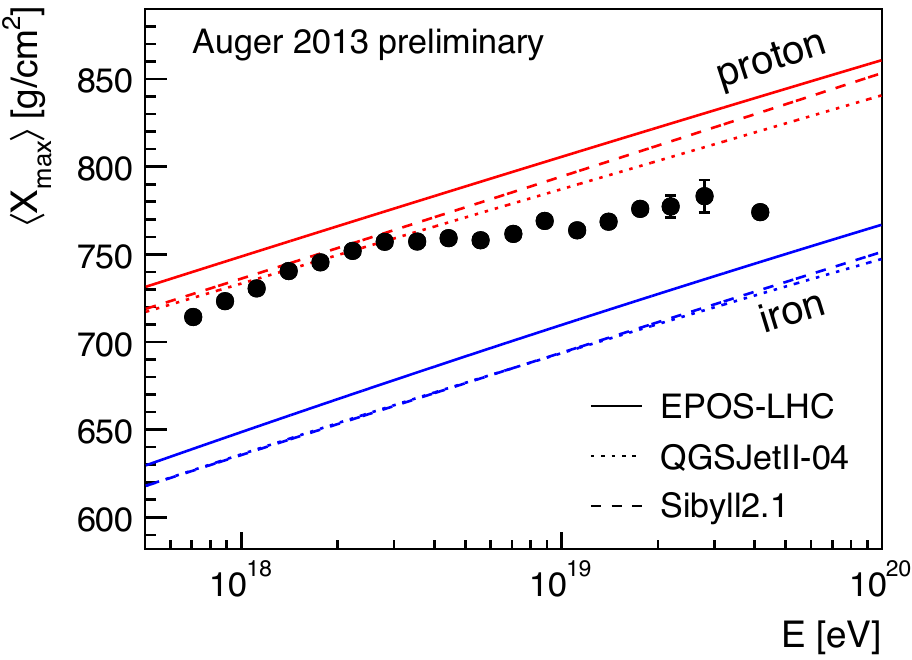}\hfill
\includegraphics[width=0.48\textwidth]{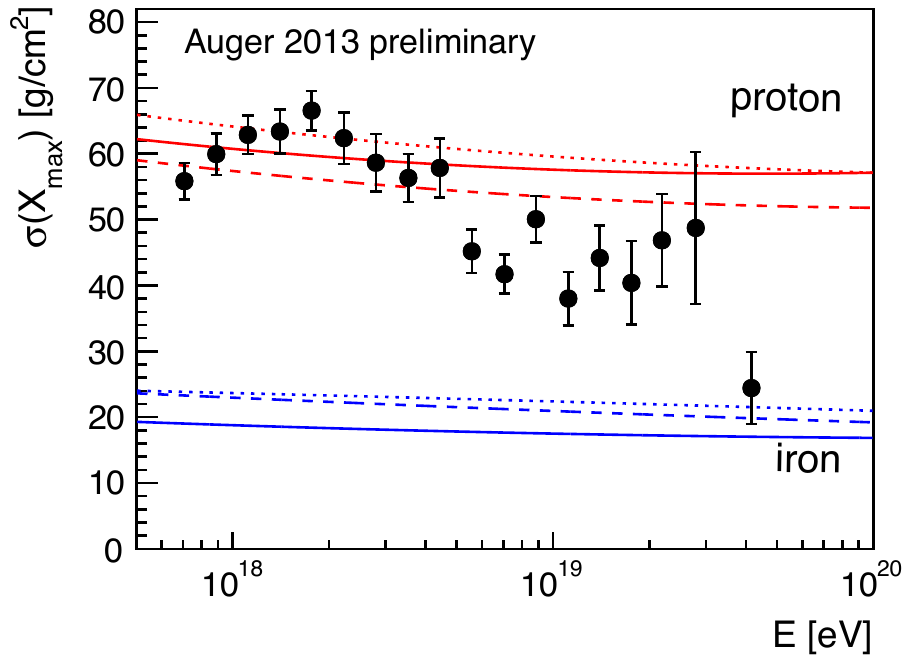}
\caption{Evolution of $\langle$\Xmax$\rangle$ and $\sigma\,_{X_{max}}$ as a function of energy. Measurements are from the hybrid data set of Auger. (\cite{AC:Sein}).}
\label{Xmax}
\end{figure*}

This unprecedented statistical accuracy allows clear identification of two features in the energy spectrum, the {\it Ankle} and a {\it cut-off} at the highest energy.  At the Ankle  the spectral index changes from -3.23$\pm$0.07 to -2.63$\pm$0.04 at a break point energy of  5~\,EeV.
Above 20 EeV the spectrum starts to deviate from a simple power law and a flux suppression (a cut-off) is observed. At $E_{50\%} = 40$\,EeV the observed spectrum is half of what is expected from the extrapolation of the power law observed just above the Ankle. When compared to a simple continuation of a power-law, the significance of the cut-off is more than 20 sigma, however its origin, as that of the Ankle is yet to be determined.   

These features can originate from interactions of the cosmic rays with the intergalactic radiation field (mainly the CMB) during their transport from their sources to the Earth. This is the case for example of the e$^+$e$^-$ pair or pion production (GZK) from protons off the CMB photons for the Ankle and the cut-off respectively or of the photo-disintegration of nuclei. Such features also can originate from the sources spatial distributions and/or their acceleration characteristics, in this case the Ankle could sign the transition from a Galactic dominated cosmic ray sky to an extra-galactic dominated one while the cut-off would directly reflect the maximum energy reachable by  the sources themselves. Various scenarios have been put forward, combining these possible origins in various ways (see e.g.~\cite{review} for an overview).   

The models shown in figure~\ref{f:comb_astromodels} assume either a pure proton or pure iron composition. The fluxes result from different assumptions of the spectral index $\beta$ of the source injection spectrum and the source cosmological evolution parameter $m$. The maximum energy of the source was set in these particular examples to 100 EeV and 300 EeV, the former describing better the data in the cut-off region. The model lines have been calculated using CRPropa~\cite{CRPropa} and validated with SimProp~\cite{SimProp}. 

Despite its high statistical accuracy, the energy spectrum alone is not sufficient to distinguish between the various scenarios. There are simply too many unknowns (source distributions and evolution, acceleration characteristics, cosmic ray mass composition). Other observables such as anisotropies and mass composition parameters will have to be combined to disentangle the situation.

\section{Mass composition}
The hybrid nature of the Auger observatory allows for a very precise measurement of the shower longitudinal profile on a subset of less than 10\% of the events (the hybrid data set). The combination of the FD and SD allows for a precise determination of the shower geometry which in turn allows measurement of the position of the maximum shower size (\Xmax) with an accuracy of better than 20\,g\,/\,cm$^2$.

The updated (but preliminary) results regarding the evolution with energy of the two first moments of the \Xmax\,distributions are shown in Fig.~\ref{Xmax}. When compared to the model lines, the data clearly indicate a change of behavior at a few EeV, i.e. in the Ankle region. 

While predictions of different models may not be an accurate representation of nature for the  absolute values of $\langle$\Xmax$\rangle$, hence making it difficult to convert with confidence this data into mass values, they have similar predictions (within 20 g/cm$^2$ for $\langle$\Xmax$\rangle$\, and 10~g/cm$^2$ for $\sigma\,_{X_{max}}$) for those parameters. In particular, all models predict that for a constant composition the elongation rate (slope of the $\langle$\Xmax$\rangle$\, evolution) and  $\sigma\,_{X_{max}}$ are also constant as a function of energy. This is at clear variance from the measurements themselves. Hence, under the hypothesis that no new interaction phenomena in the air shower development come into play in that energy range, the data clearly support that the composition evolves in the Ankle region.

\begin{figure*}[!t]
\includegraphics[width=\textwidth]{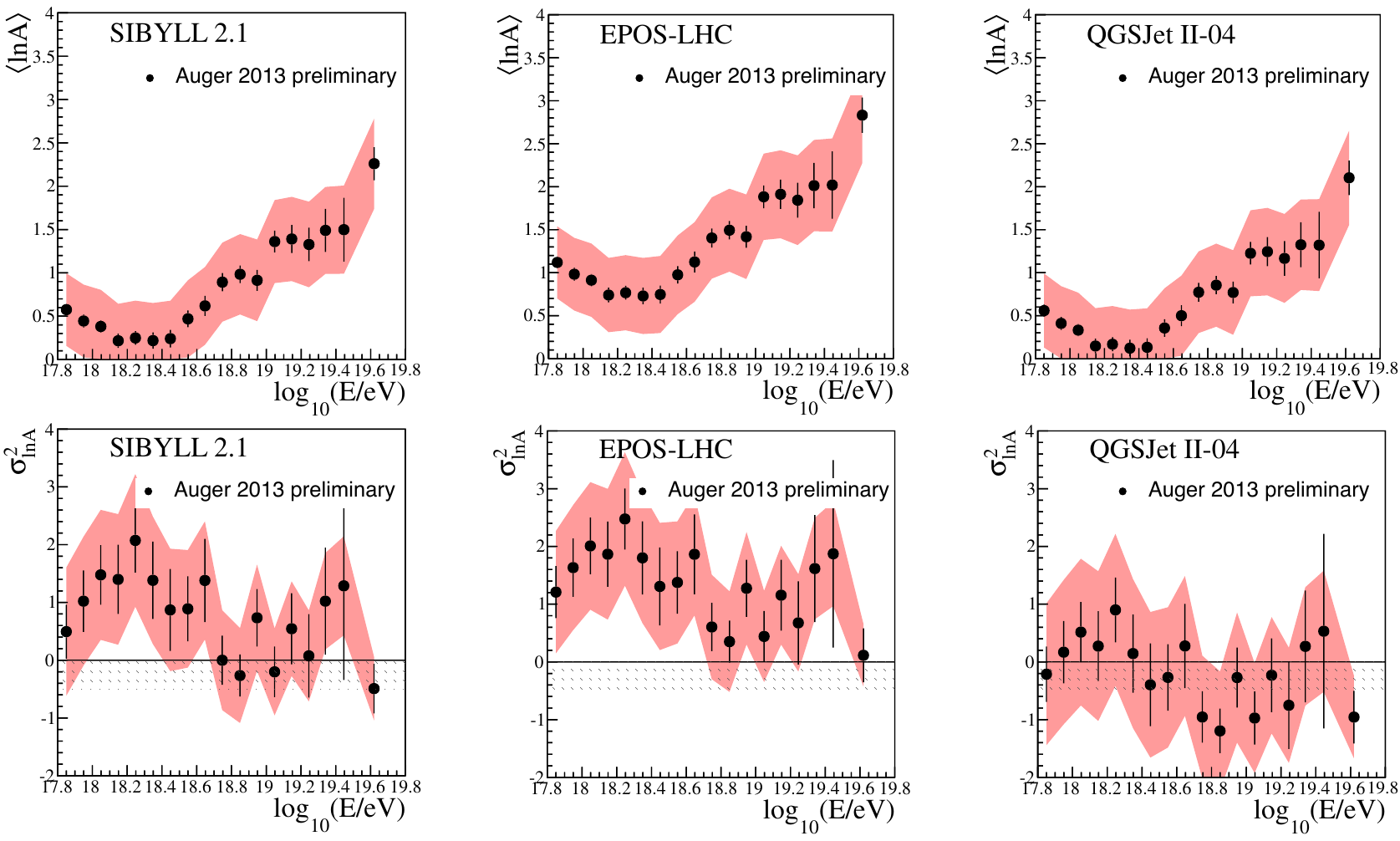}
\caption{Conversion to $\langle$lnA$\rangle$ and $\sigma^2\,_{\rm lnA}$  using various hadronic interaction models. The red bands indicate the systematic uncertainties.(\cite{AC:Sein}).}
\label{LnA}
\end{figure*}

While subject to the belief that current interaction models do represent reality, it is possible to convert the measured data into the first two moments of the lnA distribution at the top of the atmosphere~\cite{LNA}. This is shown in Fig.~\ref{LnA} using several hadronic interaction models~\cite{SIBYLL2.1,EPOS-LHC,QGSJetII-04}. From this conversion it is possible to interpret the aforementioned evolution as a change from light to medium light composition with a minimum in the average lnA just before the Ankle, i.e. between 2 and 3\,EeV. Looking at the  $\sigma^2\,_{\rm lnA}$ plot, one can also argue that the evolution is slow in terms of masses ($\sigma^2\,_{\rm lnA}$ stays below 2 in the whole range indicating that the mix is between nearby masses rather than between proton and iron)\footnote{ $\langle$lnA$\rangle$ is 0 for pure proton and 4 for pure iron while $\sigma^2\,_{\rm lnA}$ is 0 for pure composition and 4 for a 50:50 p/Fe mix.}. We also observed that for some model the central predicted variance of lnA is negative but this is not the case within our systematic uncertainties. 

\section{Hadronic Interactions}
We have performed several analyses to extract a muon size parameter from the hybrid or SD data sets. These analyses~\cite{AC:DiegoG,AC:Ines,AC:Balazs,AC:Glennys} all indicate that current hadronic interaction models predict muon sizes that are smaller (by at least 20\%) than observed in the data, unless one assumes that the data is composed  of pure iron which is in contradiction, according to the same models, with the observed \Xmax\, distributions.

In~\cite{AC:Glennys} we have selected all showers (411) measured in hybrid mode with an energy between 10$^{0.8}$ and 10$^{1.2}$\,EeV. For each of those showers, we have generated Monte Carlo events with similar energies selecting those which also matched  the measured longitudinal profile. Then, for those matching events, the predicted lateral distribution of the signal has been compared to the data recorded by the SD. 

\begin{figure*}[!t]
\begin{minipage}{0.49\textwidth}
\hspace*{-1.2cm}
\includegraphics[width=1.15\columnwidth]{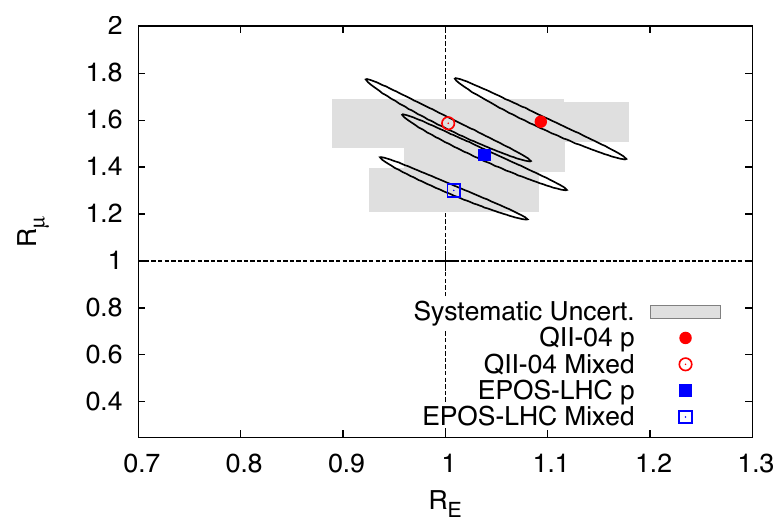}
\caption{Value of the energy rescaling parameter $R_E$ and muon rescaling parameter $R_\mu$ that best represent the Auger hybrid data at 10\,EeV. The predicted energy is compatible with the observed one (R$_E$ is compatible with 1 within the systematics on the absolute energy scale) while the muon rescaling parameters demands an increase of at least 20\% of the muon size from the models. (\cite{AC:Glennys}).}
\label{hadint}
\end{minipage}
\hfill
\begin{minipage}{0.49\textwidth}
\begin{minipage}{1.1\columnwidth}
\hspace*{-0.6cm}
\includegraphics[width=0.49\columnwidth]{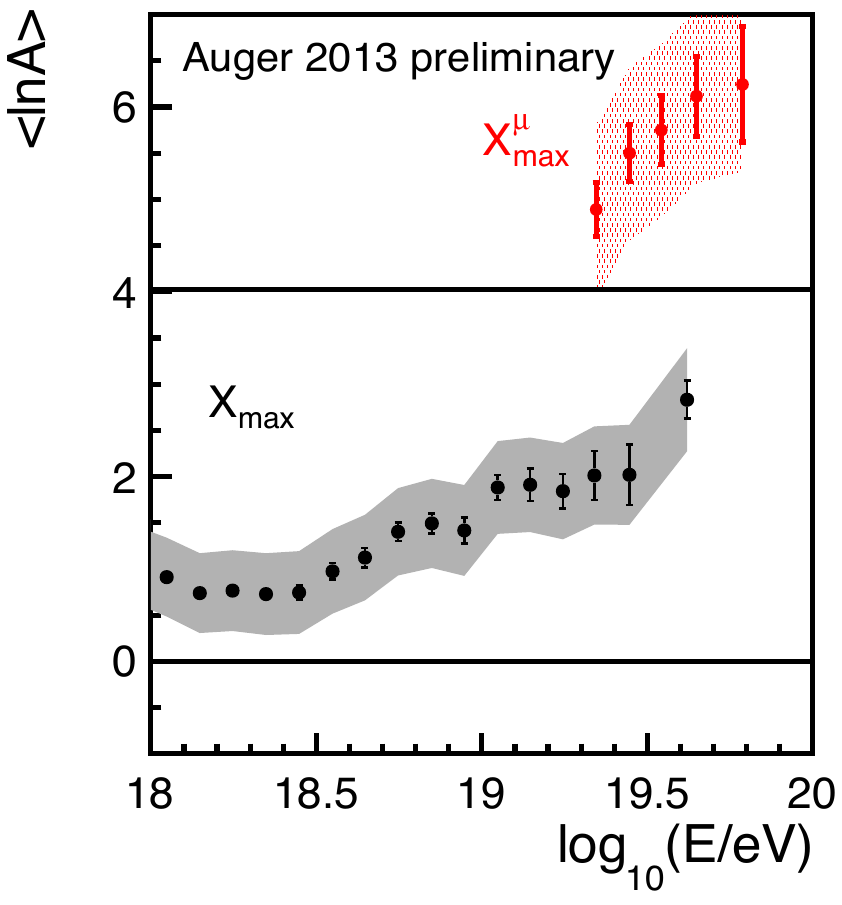}
\includegraphics[width=0.49\columnwidth]{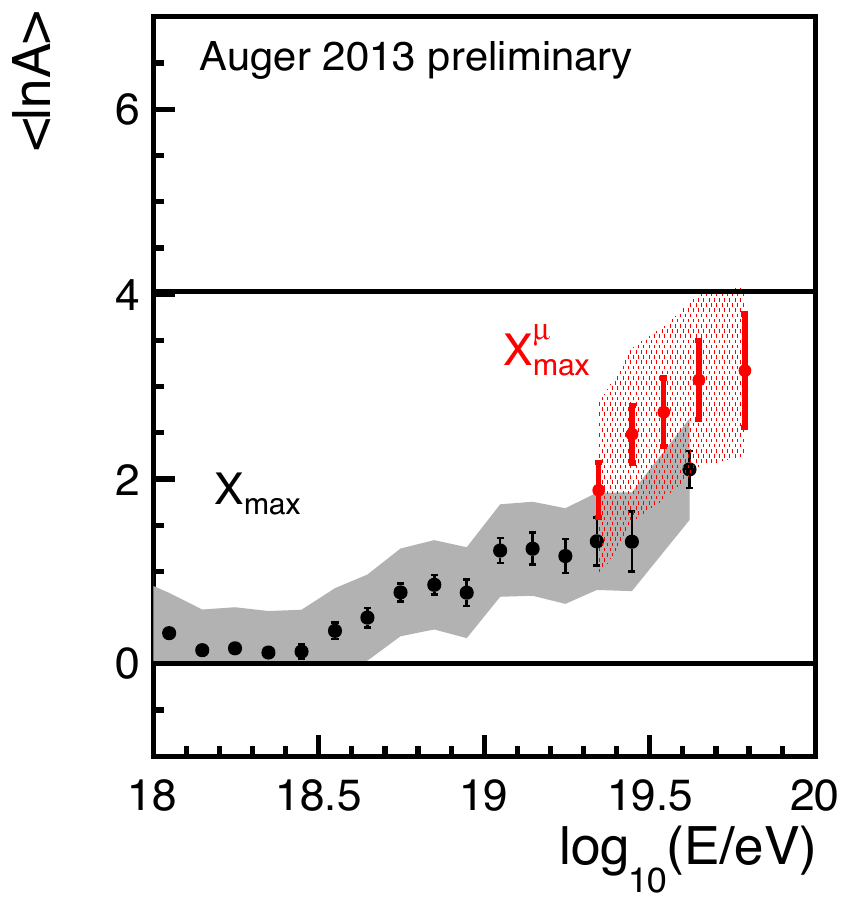}
\end{minipage}
\caption{Conversion of the \Xmax\, and  X$^\mu_{max}$ observable to $\langle$lnA$\rangle$ using two different hadronic interaction models EPOSS-LHC (left) and QGSJetII-04 (right). While QGSJetIII-04 present a more coherent conversion, EPOS-LHC offers a better description of the rapidity gap distribution of p-p collision at the LHC. The modification of this distribution in EPOS to better reproduce the LHC p-p data is believed to be responsible for the shift in X$^\mu_{max}$~\cite{Tanguy}. }
\label{Xmu}
\end{minipage}
\end{figure*}

The Monte Carlo predictions have been found to be systematically below the observed signals, regardless of the hadronic model being used. To match the lateral distributions we introduce two parameters that have been adjusted to the data. These parameters are $R_E$ which acts as a rescaling of the shower energy, and $R_\mu$ which acts as a muon size rescaling factor. The values that best reproduce the data are shown in Fig.~\ref{hadint} for a set of proton showers only and for a set showers from a mixed composition sample whose global \Xmax\, distribution matches that of the data. 

In all cases the $R_\mu$ rescaling factor is larger than one, indicating a deficit in the predictions, while for $R_E$ it is compatible with 1 for the mixed set and also for the pure proton set but only within the systematic uncertainties (mainly originating form our absolute energy scale). Independent analyses using inclined showers or relying on the distinct signal shape left by muons in the WCD also point to a deficit of muons in the simulations~\cite{AC:Ines,AC:Balazs}. 

In another study, based purely on the SD data we have reconstructed the {\it muon production depth} profile (MPD, \cite{AC:DiegoG}). From this profile it is possible to extract the depth of maximum production of the muons that reach the ground (X$^\mu_{max}$ ) which is also a mass indicator as it is linked to the longitudinal evolution of the EAS in the atmosphere. 

An interesting aspect of this study is that it gives us a second observable, similar to \Xmax, that can be converted into $\langle$lnA$\rangle$. It is therefore tempting to convert both our \Xmax\, and X$^\mu_{max}$ data
into $\langle$lnA$\rangle$ using the same interaction model. The result of such conversion is shown in Fig.~\ref{Xmu} for two models. In the first case, with EPOS-LHC, the two observables convert into an incompatible mass value. According to the model authors~\cite{Tanguy} this is linked to the better representation of  the rapidity gap distribution of pp interactions measured at the LHC. Of course, UHECR collision in atmosphere are not p-p collisions but at least  p-Air collisions if not higher masses. The observed apparent contradiction could then simply point at collective effects of the nuclei collisions in the atmosphere. The representation from the second model, QGSJetII-04, seems better but in that case the rapidity gap distribution from the model is in poorer agreement with the LHC data. While one cannot conclude on the quality of a given model from this plot alone, this analysis shows the interest and the power of UHECR data to constrain high energy interaction models.

\begin{figure*}[!t]
\center
\includegraphics[width=\columnwidth]{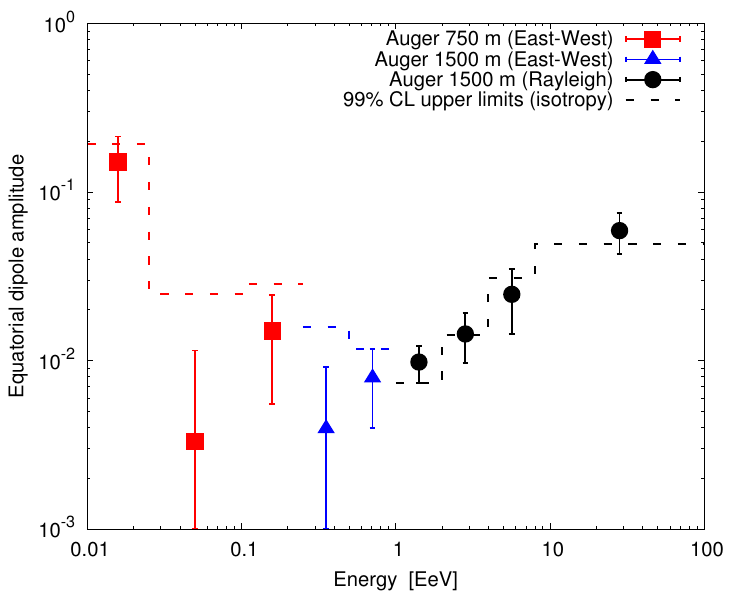}
\includegraphics[width=\columnwidth]{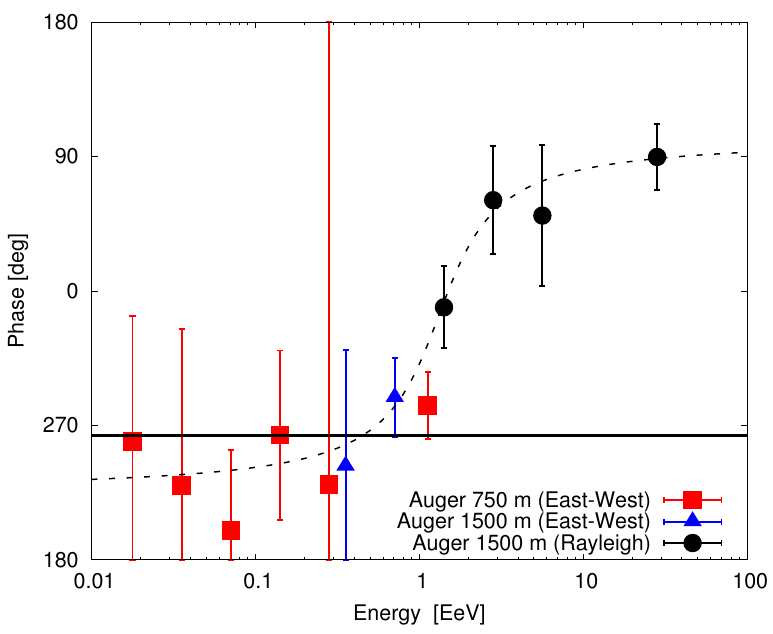}
\caption{Equatorial dipole amplitude (left) and phase (right) evolution as a function of energy. Black circle : modified Rayleigh analysis, blue triangles : East-West analysis, red squares infill data with East-West analysis. Three point lie above the 99\%CL line in the amplitude plot while the phase shows a smooth evolution from the galactic centre towards the galactic anti-centre directions. (\cite{AC:Ivan}).}
\label{isot1}
\end{figure*}

\section{Anisotropies}
The Auger collaboration has also performed extended analyses of the UHECR arrival direction distributions in several energy ranges and different angular scales~\cite{AC:Ivan,AC:Rogerio,AC:Benoit,AC:Francisco}.

Some particularly interesting results come out of the analysis of the first harmonic modulation in the right ascension distribution of the events~\cite{AC:Ivan}. The results of this analysis on the equatorial dipole amplitudes is shown in Fig~\ref{isot1} for an extended range in energy covering nearly 4 orders of magnitude. While no clear evidence for anisotropy has been found yet it is remarkable to see that in the range above 1\,EeV, 3 out of the 4 points are above the 99\% CL line, i.e. only one percent of isotropic samples would show equal or larger amplitudes. 

The phase evolution in the same energy range, also shown in Fig.~\ref{isot1}, has an interesting behavior with a smooth transition from the galactic centre direction (270$^\circ$) to 90$^\circ$. To test the hypothesis that the
phase is undergoing a smooth transition, we began to independently analyze data obtained after April 2011. After 18 months the new and independent data set is showing a similar trend~\cite{AC:Ivan}. Another 18 months of data collection to reach an aperture of 21,000\,km$^2$\,sr with the independent data set is needed before the trend can be confirmed.  

\begin{figure}[!t]
\includegraphics[width=\columnwidth]{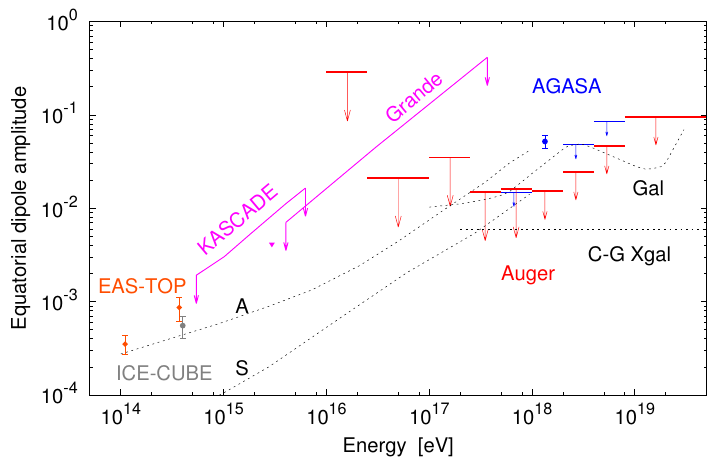}
\caption{Upper limit at 99\%CL for the equatorial dipole amplitude as a function of energy. In red are the limits obtained over the full energy range of the Auger Observatory. Results from AGASA are shown in blue, from KASCADE and KASCADE-Grande in magenta, EAS-TOP in orange and ICE-CUBE in grey. Predictions from different models are displayed, labeled as A, S, Gal and C-G Xgal (see text).(\cite{AC:Ivan}).}
\label{isot2}
\end{figure}

It is interesting to note that despite the possible hints for CR anisotropy discussed above, any such anisotropy would be remarkably small (at the \% level). The Auger collaboration is therefore able to place stringent limits on the equatorial dipole amplitude d$_\perp$ as shown in Fig.~\ref{isot2}.   In this figure, the predictions labeled A and S correspond to models in which cosmic rays at 1 EeV are predominantly of galactic origin. They escape from the galaxy by diffusion and drift motion and this causes the predicted  anisotropies. A and S stand for two different galactic magnetic field symmetries (antisymmetric and symmetric). In the model labeled Gal~\cite{Calvez} a purely galactic origin is assumed for all cosmic rays up to the highest energies. In this case the anisotropy is caused by purely diffusive motion due to the turbulent component of the magnetic field. Some of these amplitudes are challenged by our current bounds. The prediction labeled C-G Xgal is the expectation from the Compton-Getting effect for extragalactic cosmic rays due to the motion of our galaxy with respect to the frame of extragalactic isotropy, assumed to be determined by the cosmic microwave background. 

The bounds reported here already exclude the particular model with an antisymmetric halo magnetic field (A) above energies of 0.25 EeV and the Gal model at few EeV energies, and are starting to become sensitive to the predictions of the model with a symmetric field. (see~\cite{AC:Ivan} and references therein for more details).

\begin{figure}[!t]
\includegraphics[width=\columnwidth]{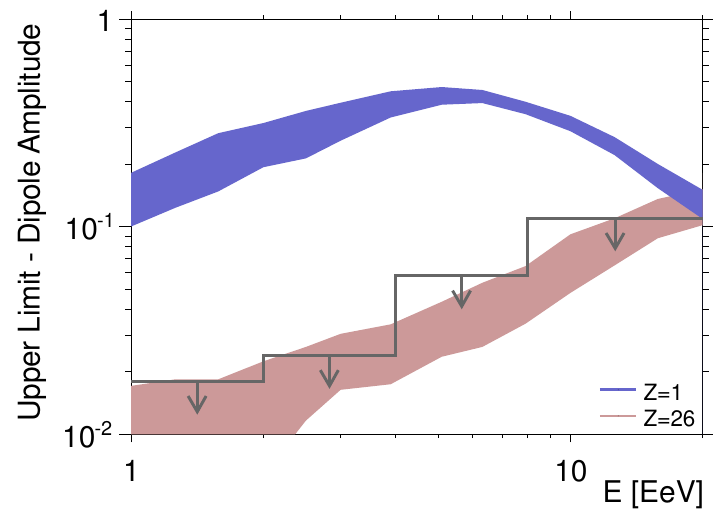}
\caption{Upper limit at 99\%CL for the  dipole amplitude as a function of energy. Some generic anisotropy expectations from stationary galactic sources distributed in the disk are also shown, for various assumptions on the cosmic ray composition. (\cite{AC:Rogerio}).}
\label{isot3}
\end{figure}

We have also conducted searches for dipole and quadrupole modulations reconstructed simultaneously in declination and right ascension. The upper limits presented in~\cite{AC:Rogerio} are shown in Fig.~\ref{isot3}. They are presented along with generic estimates of the dipole amplitudes expected from stationary galactic sources distributed in the disk considering two extreme cases of single primaries: protons and iron nuclei. This figure illustrates the potential power of these observational limits. 

While other magnetic field models, source distributions and emission assumptions must be considered, in this particular examples we can exclude the hypothesis that the light component of cosmic rays comes from stationary sources densely distributed in the Galactic disk and emitting in all directions.

\begin{figure*}[!t]
\center
\includegraphics[width=\textwidth]{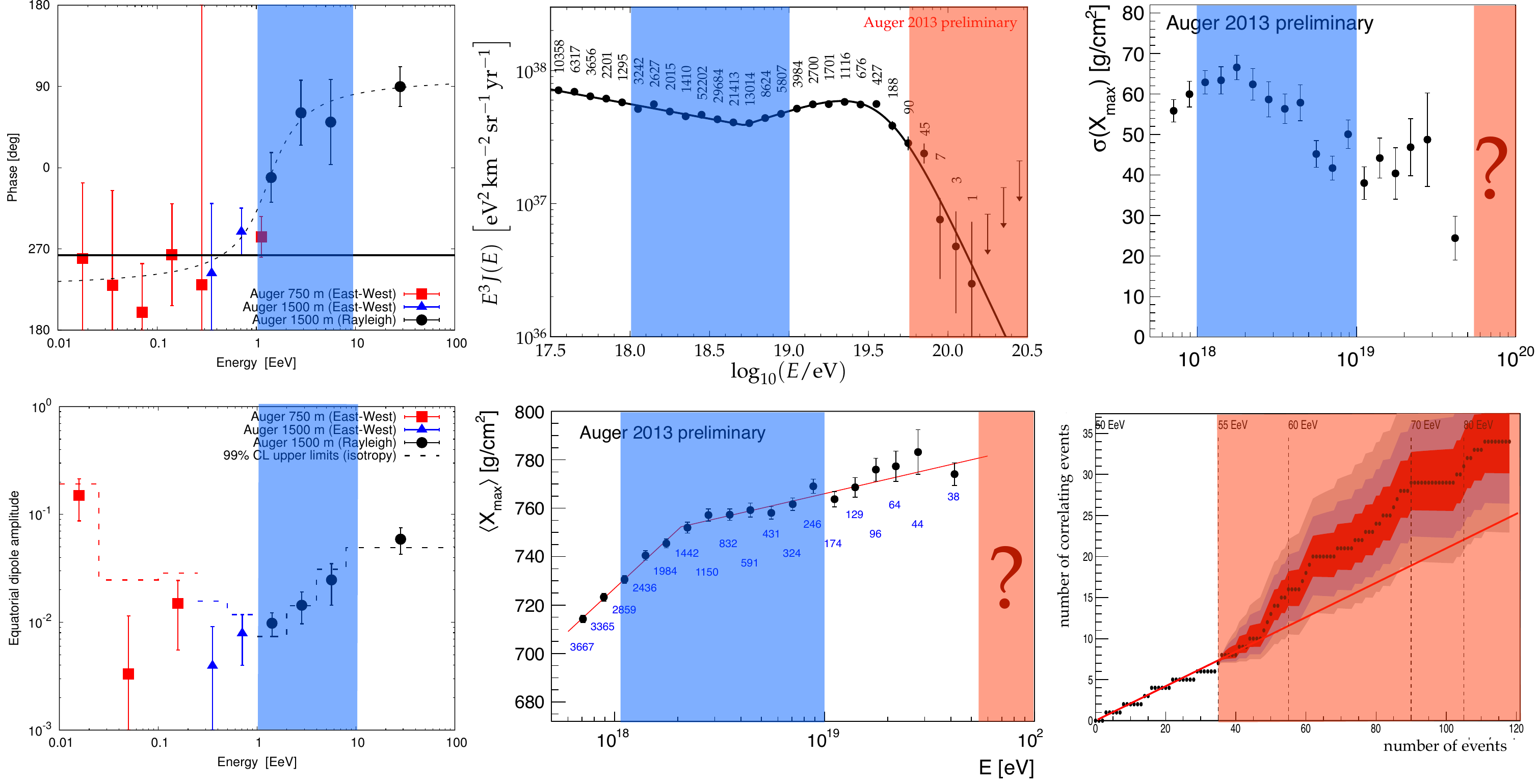}
\caption{An overall view of the Auger results showing the variety of the observables and the coherence of their behavior. The blue bands correspond to the Ankle region where features are observed in the spectrum, mass and anisotropy data. The red bands corresponds to the cut-off region where, unfortunately, due to the low duty cycle of the fluorescence technique the mass information is missing. For completeness the VCV correlation (from~\cite{Kampert2011}) is also shown as an energy ordered plot. The onset of the correlation signal is visible at about 55 EeV.}
\label{conclusion}
\end{figure*}

\section{Conclusions}
The Auger observatory is producing measurements of the UHECR properties over 4 orders of magnitude in energy (from 0.01\,Eev to above 100\,EeV). A synthesis of those measurements is presented in Fig.~\ref{conclusion} where one can scrutinize the  quality and coherence of those observations. 

The astrophysical interpretation of that data is however still delicate as most properties of the UHECR sources are still unknown. When treating the sources distributions and cosmological evolutions, their spectral indexes, their compositions and their maximum energies as free parameters many different interpretations can lead to an acceptable reproduction of our \Xmax\, spectrum data. Leaving alone the fact that all sources need not to be equal ! Additionally, the inclusion of our anisotropy results adds more complexity but, there again, the unknowns on the Galactic and extragalactic magnetic fields and on the source distributions and composition leave much space for speculations. 

Nevertheless, taking at face value the current model conversion of our \Xmax\,data  into masses and adding the information of our spectrum measurement, it is possible that the cut-off region represents more a consequence of the source maximal acceleration energy (of the order of 4\, EeV for proton) than a propagation effect as expected from the GZK scenario. However, taking into account the remaining non-trivial correlation observed in our highest energy events with the VCV catalog (see figure~\ref{conclusion}, the correlation signal is  2$\sigma$ above the expected fraction for an isotropic sky) the presence of a sub-dominant fraction (less than about 20\%) of protons may be expected in this region. The identification of this sub-dominant fraction will require an excellent mass determination capability in this energy range - something similar to the current FD performances on the measurement of the EAS longitudinal development but with a 100\% duty cycle. Note also that in such scenarios the spectral features originate from the sources properties rather than from interaction of the  bulk of the cosmic rays with the CMB.  Magnetic deflections in transit to Earth also are important.

Still in the cut-off region another interpretative option is to consider a possible change in the hadronic interactions of protons at the highest energies. Such modification would make the proton EAS look like those currently modeled from heavier nuclei. The difficulty encountered in constraining the high energy interaction generators at energies one or two orders of magnitude above the LHC leaves some room for such a scenario. Additional data from UHECR including in particular the muonic content of EAS will definitely help in reducing those unknowns.    

In the Ankle region the question is still open as to whether the break observed in the spectrum is the consequence of a propagation effect or the signature of a transition between two types of  sources (be they both Galactic or not). Several key observables, if they are combined,  will help to resolve the issue. An anisotropy study for at least two different mass spectra (one light, one heavy)  from 0.1\,EeV up to 10\,EeV would for example allow to distinguish between a propagation effect and a source transition scenario. The key is to cover a wide enough energy range to connect adequately the new data to that measured by observatories at lower energies such as those from KASCADE-Grande~\cite{KASCADE}.

Additional information such as the limits on the photon fractions in the EeV range and/or the neutrino fluxes will also bring interesting light into both regions. The absence of cosmogenic photons or neutrinos, for example, would indicate clearly that there are no (or very few) proton sources in the cosmos with limiting energy well above the GZK cut-off. 

The Auger observatory will continue taking data for the years to come and the collaboration is deeply engaged in improvements and upgrades of our detection systems. We aim at covering the open issues discussed above. 

At the low energy end (between 0.01 and 1\,EeV) we have the HEAT and AMIGA extensions. We have also recently modified the local trigger conditions of the surface array detectors to lower our full trigger efficiency threshold. It is now about 1 EeV  for the 1.5\,km  array (it was 3\,EeV before). This improvement will provide us with about 5 times more events in this energy range than what we had before. This will  allow us to augment significantly our sensitivity to anisotropy searches. In addition, because this new triggering scheme is less sensitive to individual muons entering the WCDs,  it will allow us to improve significantly our photon sensitivity. Together with the increased statistics this opens great perspectives for the cosmogenic photon searches. 

At the high energy end, the upgrade of our SD array is under study to provide us with a detector able to measure both the muon content and the age of the shower at ground. This two observables will give us the means to identify the UHECR composition on an event-by-event basis up to the highest energies. The collaboration is evaluating several detector options that can in principle fulfill these ambitious scientific goals~\cite{Proposal}.

\vspace*{0.5cm}
\noindent \footnotesize{{\bf Acknowledgment:}{The successful installation, commissioning, and operation of the Pierre Auger Observatory would not have been possible without the technical and administrative staff in Malargüe.
We acknowledge their extraordinary commitment to this scientific endeavor. ALS acknowledges financial supports by the French National Institute for nuclear and particle physics  (CNRS/IN2P3) and by the University Pierre \& Marie Curie in Paris (UPMC).}}

\end{document}